\documentclass[a4paper,11pt]{article}
\usepackage{jheppub} 
\usepackage{lineno}

\usepackage[T1]{fontenc} 
\usepackage{xcolor}
\usepackage{slashed}
\usepackage{subcaption}
\usepackage{physics}
\usepackage{csquotes}
\usepackage{caption}
\usepackage{natbib}
\usepackage{amsmath}
\usepackage{amssymb}
\usepackage{bm}
\usepackage{bbm}
\usepackage{hyperref}
\usepackage{dsfont}
\usepackage{todonotes}
\usepackage{siunitx}

\newcommand{\ignore}[1]{}

\newcommand{\R}{\mathbb{R}}
\newcommand{\meff}{m_{\tilde{\chi}, \text{eff}}}
\newcommand{\mgrav}{ \ensuremath{m_{3/2}} }
\newcommand{\mpl}{\ensuremath{m_\text{pl}}}

\title{\boldmath Gravitational Dark Matter Production in Supergravity $\alpha$-Attractor Inflation}

\author[a]{Chenhuan Wang}
\author[a]{Wenbin Zhao}
\affiliation[a]{Bethe Center for Theoretical Physics and Physikalisches Institut, Universität Bonn,\\ \indent
Nussallee 12, 53115 Bonn, Germany}

\emailAdd{cwang1@uni-bonn.de}
\emailAdd{wenbin.zhao@uni-bonn.de}
\date{\today}

\abstract{We consider gravitational particle production (GPP) of dark matter (DM) under a supergravity framework, where the $\alpha$-attractor inflation model is used. The particle spectrum is computed numerically and the DM number density is obtained. We show how the DM mass, gravitino mass and inflation model parameters modify the results, and find the reheating temperature which leads to sufficient DM production. In our setup, supergravity corrections suppress the efficiency of GPP, and make the isocurvature constraint much weaker compared with the normal case. With tensor-to-scalar ratio ranging from $10^{-3}-10^{-4}$ and DM mass from $10^{-2} m_\phi - m_\phi$, the required reheating temperature should be around $10^3 \textrm{GeV} - 10^7 \textrm{GeV}$.}

\begin{document}
\maketitle
\flushbottom
\section{Introduction}
Based on current observations, on large scale our Universe is almost isotropic and presumably also homogeneous. It contains dark matter (DM) that interacts weakly with the ordinary matter~\cite{Planck:2018vyg}. Inflation, where the Universe experiences a period of accelerated expansion, provides an excellent explanation for this isotropy and homogeneity~\cite{Linde:1981mu, Albrecht:1982wi}. The quantum fluctuations of the inflaton field also seed the initial conditions for structure formation. In addition, inflation can source particle production; one example is the so-called gravitational particle production (GPP), due to the time evolution of the background spacetime~\cite{Lyth:1996yj,Chung:1998zb}.

GPP is inevitable for any degree of freedom as long as it couples to gravity. This production channel can contribute significantly to the DM abundance, and can generate subsequently DM isocurvature perturbations. For a comprehensive review, we refer to~\cite{Kolb:2023ydq}. If gravity is the only interaction between DM and the Standard Model (SM) particles, DM from GPP will never be in thermal equilibrium with the rest of particles, unlike canonical thermal freeze-out mechanism. GPP has been extensively studied in the past decades, leading to a rich phenomenology for scalar and higher spin fields~\cite{Lyth:1996yj, Chung:1998ua,Kolb:1998ki,Mukhanov:2005sc,Ling:2021zlj,Garcia:2023qab,Kuzmin:1998kk,Chung:2011ck,Graham:2015rva,Ahmed:2020fhc,Kolb:2020fwh,Gorghetto:2022sue,Cembranos:2023qph,Ozsoy:2023gnl,Hasegawa:2017hgd,Kallosh:1999jj,Antoniadis:2021jtg,Kaneta:2023uwi,Casagrande:2023fjk,Kolb:2023dzp,Nilles:2001ry,Nilles:2001fg}. Meanwhile, analytic treatments to track particle production have been developed in certain cases~\cite{Chung:1998bt, Ema:2018ucl, Cembranos:2019qlm, Enomoto:2020xlf, Basso:2021whd,Hashiba:2021npn,Basso:2022tpd,Kaneta:2022gug,Racco:2024aac,Verner:2024agh,Jenks:2024fiu}.

Even though the precise results of GPP depend on the inflation model, some general statements can be made without specifying inflation dynamics. For a scalar field $\chi$ lighter than the inflation scale $H_I$, its number density spectrum will feature an infra-red (IR) divergence, which has to be regulated through a cut-off in momentum space. Its isocurvature spectrum will be almost scale invariant, which might conflict with CMB observations. In contrast, if the field is heavier than $H_I$, its number density will be convergent without cut-off and the isocurvature signal can be negligible.

Since gravitational interactions at sub-Planckian energies are weak, the above conclusion can be easily modified. Corrections may appear from various sources. Introducing a non-minimal coupling between DM and gravity can greatly change the efficiency of GPP. A conformally coupled scalar experiences less GPP, has no IR divergence and leaves negligible isocurvature perturbations at CMB scales, larger interaction will enhance the GPP production~\cite{Markkanen:2015xuw, Fairbairn:2018bsw, Cembranos:2019qlm, Clery:2022wib, Garcia:2023qab, Yu:2023ity}. Introducing a direct coupling between the inflaton field and DM leads to a similar effects~\cite{Garcia:2022vwm,Garcia:2023awt}. In this paper we would like to focus on another possibility, where such corrections comes from supergravity (SUGRA) effects.

SUGRA is a local gauge theory of supersymmetry (SUSY), which generally predicts non-renormalizable interactions between the inflaton field and DM. These interactions can play a significant role in cosmology~\cite{Endo:2006qk,Endo:2006nj,Endo:2007sz,Enomoto:2013mla,Ema:2024sit}. Previous study of GPP in SUGRA setting focuses on the case where $m_\chi> H_I$~\cite{Nakayama:2019yjv}, and we would like to have a comprehensive understanding of the entire parameter space. Even for light DM, SUGRA correction lifts its effective mass naturally higher than Hubble scale. This correction has the same origin as the $\eta$ problem in SUGRA inflation research~\cite{Copeland:1994vg}.

In this paper, we focus on the $\alpha$-attractor model, which is realized in SUGRA using a stabilizer field~\cite{DallAgata:2014qsj,Kallosh:2014hxa}. This framework also allows us to introduce an arbitrary SUSY breaking scale $m_{3/2}$. 
Even though for the canonical choice of $m_{3/2}\sim \SI{1}{\tera\electronvolt}$, its impact on GPP can be safely neglected, it is still interesting to investigate how a larger $m_{3/2}$ could change GPP. 
We numerically calculate the number density of a scalar DM, as well as its isocurvature contribution. Compared with non-SUSY GPP, SUGRA models need a higher reheating temperature to produce the same amount of DM.

In this work, we use the reduced Planck mass $\mpl = (8 \pi G)^{-1/2} = \SI{2.44e18}{\giga\electronvolt}$, except in section \ref{sec:SUGRA} where the Planck units are used, i.e. $\mpl = 1$. This paper is organized as follows. In section \ref{sec:theory}, we introduce the model of inflation, the basics of GPP, and show how to embed our model in a SUGRA framework. Numerical results for GPP are presented in section \ref{sec:num} and its phenomenological consequences are discussed. Section \ref{sec:conc} contains our conclusion and outlook of this work. We give a short derivation of DM relic abundance in Appendix~\ref{sec:app_rho_s0}.

\section{Theoretical framework}
\label{sec:theory}
\subsection{Inflation model}
Inflation is assumed to be a period of quasi-de Sitter spacetime. The simplest realization is single-field slow-roll inflation, where the potential energy dominates the energy density of the scalar field. This leads the Universe to expand (approximately) exponentially. 
As the inflaton field is nearly homogeneous, we can use the Friedmann-Lemaître-Robertson-Walker (FLRW) metric to describe the background geometry
\begin{equation}
    g_{\mu\nu} \dd{x^\mu} \dd{x^\nu} = a^2(\tau) \left( \dd{\tau}^2 - \dd{\vec{x}^2} \right)\,,
    \label{eq:conf-metric}
\end{equation}
where $\tau$ is the conformal time and $a(\tau)$ is the scale factor. The cosmic time $t$ satisfies $dt = a d\tau$. The equation of motion of the inflaton field and the Friedmann equations of first and second kind are
\begin{subequations}
\begin{align}
    \phi '' &+ 2 a \mathcal{H} \phi' + a^2 \pdv{V}{\phi} = 0\,, \\
    \mathcal{H}^2 &= \left( \frac{a'}{a} \right)^2 = \frac{1}{3\mpl^2} \left( \frac{\phi'^2}{2a^2} + V \right)\,,  \\
    \frac{a''}{a} &= \frac{1}{6 \mpl^2} \left(4a^2 V - \phi'^2 \right)\,,
\end{align}
\label{math:background-eq}
\end{subequations}
where the $'$ denotes the derivative with respective to conformal time $\tau$ and $\mathcal{H}$ is the conformal Hubble parameter. The slow-roll parameters are given by the first and second derivative of the inflaton potential
\begin{equation}
    \epsilon_V = \frac{\mpl^2}{2 } \left( \frac{V_\phi}{V}\right)^2\,, \quad \eta_V = \mpl^2 \frac{V_{\phi\phi}}{V}\,.
\end{equation}
where the subscript $\phi$ denotes the derivative with respect to $\phi$ field, i.e. $V_\phi = d V(\phi) / d \phi $. Slow-roll inflation occurs when both $\epsilon_V, |\eta_V| \ll 1$.

In this work, we investigate the possibility that T-mode $\alpha$-attractor is the inflation model, which can be easily realized in the SUGRA context, see section \ref{sec:SUGRA}. It has the following potential \cite{Kallosh:2013yoa}
\begin{equation}
    V_I(\phi)= V_0 \tanh^{2n}{\left(\frac{\phi}{\sqrt{6\alpha}}\right)}\,.
    \label{eq:t-mode-potential}
\end{equation}
Here, the potential parameter $\alpha$ has a mass dimension 2 and controls the inflationary predictions of the model. $V_0$ denotes a energy scale. We use $n=1$ throughout the paper. The slow roll parameters for this model are
\begin{align}
\begin{split}
    \epsilon_V &= \frac{\mpl^2}{6\alpha}\frac{1}{\sinh^2(\phi/\sqrt{6\alpha}) \cosh^2(\phi / \sqrt{6\alpha})}\,,  \\
    \eta_V &= 
    \frac{\mpl^2}{6\alpha} \frac{2(1-\tanh^2(\phi/\sqrt{6\alpha})) - 6\tanh^2(\phi/\sqrt{6\alpha}) (1-\tanh^2(\phi/\sqrt{6\alpha}))}{\tanh^2(\phi/\sqrt{6\alpha})}\,.
\end{split}
\end{align}
The spectral index of the scalar perturbation $n_s$ can be given by the two slow-roll parameters at the CMB scale
\begin{equation}
    n_s = 1 + 2 \eta_* - 4 \epsilon_*\,.
\end{equation}
Here we denote quantities at such scale with subscript ${}_*$. Any viable inflation model has to give a consistent prediction on $n_s$ with the observed value, $ n_s=0.965 \pm 0.004$ by Planck 2018~\cite{Planck:2018vyg}. One can use $n_s$ to compute the field value $\phi_*$ at the CMB scale~\cite{Germ_n_2021}
\begin{equation}
    \cosh^2 \left(\frac{\phi_*}{\sqrt{6\alpha}}\right) = \frac{1}{2(1-n_s)} \left(1-n_s + \frac{4}{3} \frac{\mpl^2}{\alpha} + \sqrt{(1-n_s)^2 + \frac{8}{3} \frac{\mpl^2}{\alpha}(1-n_s) + \frac{16}{9}\frac{\mpl^4}{\alpha^2} } \right)\,.
\end{equation}
The potential parameter $\alpha$ can be expressed in terms of the tensor-to-scalar ratio $r$ and the spectral index $n_s$
\begin{equation}
    \alpha = \frac{64}{3} \frac{r}{(8(1-n_s)-r)^2 - r^2} \mpl^2 \,.
\end{equation}
Currently, the BICEP/Keck 2018 results only impose an upper limit on $r$~\cite{BICEP:2021xfz}
\begin{equation}  \label{BK2018}
r < 0.036\,\quad \text{at 95\% \text{C.L.}}\,.
\end{equation} 
This can be transferred into an upper bound on $\alpha<20 \,\mpl^2$. At last, the energy scale $V_0$ can be determined by the normalization of the curvature perturbation power spectrum $\mathcal{A}$~\cite{Planck:2018jri}
\begin{equation}
    \mathcal{A} = (2.1 \pm 0.1) \cdot 10^{-9}= \frac{1}{24\pi^2} \left.\frac{V}{\epsilon_V}\right|_{\phi=\phi_*}\,.
\end{equation}
There are two important energy scales associated with inflation, the first one is the Hubble scale during inflation
\begin{equation}
    H_\text{inf} \approx \sqrt \frac{V_0}{3\mpl^2}  \,.
\end{equation}
The second one is the inflaton mass $m_\phi$, which reads
\begin{equation}
    m_\phi^2 = \left. \pdv[2]{V}{\phi} \right|_{\phi=0} = \frac{V_0}{3\alpha}\,.
    \label{eq:Tphimass}
\end{equation}
The ratio between the inflation Hubble scale and inflaton mass $ H_\text{inf}^2 / m_\phi^2 = \alpha /\mpl^2$ depends on the potential parameter $\alpha$.

The end of slow-roll inflation is marked by either one of the magnitude of slow roll parameters being unity. The field value $\phi_e$ at the end of inflation can be obtained through
\begin{equation}
    2\cosh^2 \left(\frac{\phi_e}{\sqrt{6 \alpha}} \right) = 
    \begin{cases}
        1 + \frac{2}{3\alpha} + \sqrt{1 + \frac{8}{3\alpha}\left( \frac{1}{6\alpha} - 1 \right)}\,, & \alpha \leq \alpha_l \\
        1 + \sqrt{1 + \frac{4}{3\alpha}} \,, & \alpha > \alpha_l
    \end{cases}\,,
    \label{eq:end_of_tmode_inflation}
\end{equation}
with $\alpha_l \approx 0.17 \mpl^2 $.

\subsection{Gravitational Particle Production}
In addition to the inflaton field, we also assume there exists a spectator scalar field $\chi$ with the following action
\begin{equation}
    S_\chi = - \frac{1}{2} \int \dd[4]{x} \sqrt{-g} 
    \left( 2\xi \chi^2 R + g^{\mu\nu} \partial_\mu \chi \partial_\nu \chi + m_\chi^2 \chi^2 \right),
\end{equation}
where $g$ is the determinant of $g^{\mu\nu}$ and $R$ is the Ricci scalar. Higher order self-interactions of $\chi$ have been neglected and $m_\chi^2$ could be a function of other fields.
Here, we use the FLRW metric with conformal time defined in equation \eqref{eq:conf-metric}.

We can rescale the field $\tilde{\chi} (\vec{r}) = a (\tau) \chi (\vec{r})$. Then, after Fourier decomposing the field $\tilde{\chi}$ through comoving momentum $\vec{k}$
\begin{equation}
    \tilde{\chi} (\vec{k})= \int \frac{ d^3 \vec{r}}{(2\pi)^3} e^{i \vec{k}\cdot \vec{r}} \tilde{\chi} (\vec{r}) \,,
\end{equation}
its equation of motion turns into the equation for the mode function $\tilde{\chi}_k = \tilde{\chi} (k)$
\begin{equation}
    \tilde{\chi}_k '' + \omega^2_k \tilde{\chi}_k = 0\,, \quad \omega_k^2 = k^2 + a^2 m_{\chi, \text{eff}}^2 = k^2 + a^2 m_{\chi}^2 - (1-6\xi) \frac{a''}{a}\, ,
    \label{eq:mode-eq}
\end{equation}
where $k = |\vec{k}|$ and spatial isotropy is assumed. In this work, we only consider vanishing coupling (to gravity) $\xi = 0$. When $\omega^2_k$ varies non-adiabatically with time, the field $\chi$ can get excited and corresponding particles are produced.

The same phenomenon can also be understood in terms of the Bogoliubov coefficients. Assume that the spacetime is asymptotically flat at early and late times. The Bogoliubov coefficients describes how the orthonormal basis functions at early and late times relate to each other. The equation \eqref{eq:mode-eq} can be formulated in terms of the Bogoliubov coefficients $\alpha_k$ and $\beta_k$
\begin{equation}\label{eq:bogoliubov}
    \alpha'_k v_k = \frac{\omega'_k}{2\omega_k} v_k^* \beta_k\,, \quad \beta'_k v_k^* = \frac{\omega'_k}{2\omega_k} v_k \alpha_k\,,
\end{equation}
with
\begin{equation}
    v_k(\tau) = \frac{1}{\sqrt{2\omega_k}} \exp(-i \int^\tau \omega_k \dd{\tau'})\,.
\end{equation}
We note it is more numerically efficient to solve equations \eqref{eq:bogoliubov} instead of equations \eqref{eq:mode-eq}, except in the presence of tachyonic instability.

The comoving number density spectrum of $\chi$ field is given by
\begin{equation}
  f_\chi(k,\tau) = |\beta_k|^2\,,
\end{equation}
and the comoving number density can be obtained from it by
\begin{equation}
    {a}^3 n_\chi = \frac{1}{2\pi^2} \int \dd{k}k^2 f_\chi(k,\tau)\,.
    \label{eq:number-density}
\end{equation}
We assume that the comoving number density doesn't change after production and $\chi$ field behaves as the cold DM. The comoving energy density can be approximated as 
\begin{equation}
   a^4 \rho_\chi \approx \frac{1}{ 2 \pi^2} \int \dd k k^2 |\beta_k|^2 \omega_k\,,
\end{equation}
where we only include terms up to second order in $|\beta_k|$ and ignore the terms suppressed by the Hubble parameter and Ricci scalar (which decay away quickly).

The produced $\chi$ field also generates isocurvature perturbation. The isocurvature power spectrum is defined as 
\begin{equation}
    \Delta_\delta^2(k) = \frac{1}{\rho^2} \frac{k^3}{2\pi^2} \int \dd[3]{\vec{r}} e^{-i \vec{k}\cdot \vec{r}} \expval{\delta \hat{\rho} (0) \delta\hat{\rho}( \vec{r}) },
\end{equation}
where we have decomposed the energy density $\rho(\vec{r})= \bar{\rho}+\delta \hat{\rho}(\vec{r})$. Their time-dependencies have been omitted. The isocurvature power spectrum can be written in terms of the mode functions \cite{Chung_2005, Ling:2021zlj, Kolb:2023ydq}
\newcommand{\kPrimek}{{|\vec{k}'-\vec{k} |}}
\begin{align}
\begin{split}
    \Delta_\delta^2(\eta, k) &= \frac{1}{a^8 \bar{\rho}^2} \frac{k^3}{(2\pi)^5} \int \dd[3]{k'} \Big[ |\partial\chi_{k'}|^2 |\partial\chi_\kPrimek|^2 + a^4 m^4 |\chi_{k'}|^2 |\chi_\kPrimek|^2 \\
                            & \quad  + a^2m^2 \left( \chi_{k'} \partial\chi_{k'}^* \chi_\kPrimek \partial\chi_\kPrimek^* + \chi_{k'}^* \partial\chi_{k'} \chi_\kPrimek^* \partial\chi_\kPrimek \right) \Big]\,.
\end{split}
\end{align}
as well as the Bogoliubov coefficients with momenta relabeled through $(k, k', \kPrimek) \rightarrow (k, p, q)$ \cite{Garcia:2023awt}
\begin{equation}
    \Delta_\delta^2(k) = \frac{a_e^2 m_\chi^2}{ (a_e^4 \rho)^2} \frac{2k^2}{(2\pi)^4} \int_0^\infty \dd{p} p \int_{|k-p|}^{k+p} \dd{q} q \left[ |\beta_p|^2 |\alpha_q|^2 + \Re{\alpha_p^* \beta_p \alpha_q \beta_q^* e^{2i\Phi_p - 2i \Phi_q}}   \right].
    \label{eq:iso_formula}
\end{equation}

\subsection{SUGRA Embedment}
\label{sec:SUGRA}
In this section we discuss how to realize inflation in a SUGRA framework. In SUGRA, the scalar potential is generated by a K\"ahler potential $\mathcal{K}$ and a superpotential $\mathcal{W}$. Both of them are functions of chiral superfields $\Phi^\alpha$, where $\alpha$ refers to different superfields in the theory. Note that the K\"ahler potential $\mathcal{K}$ is a hermitian function while the superpotential $\mathcal{W}$ is a holomorphic function. In the scalar part, the K\"ahler potential determines the kinetic function of scalar fields and controls the scalar potential together with the superpotential 
\begin{equation}
\begin{split}
\sqrt{-g}\mathcal{L} &= -\mathcal{K}_{\alpha\overline{\beta}}\partial_\mu \phi^\alpha \partial^\mu \Bar{\phi}^{\overline{\beta}} - V \,,\\
V &= e^{{\cal K}}({\cal K}^{\alpha\overline{\beta}}D_{\alpha}{\cal W}\overline{D_{\beta}{\cal W}} - 3|{\cal W}|^2)\,,
\end{split}\label{eq:sugrapotential}
\end{equation}
where the Planck unit is used $\mpl = 1$ throughout this section. $\phi^\alpha$ is the spin-0 component of superfield $\Phi^\alpha$. $\mathcal{K}_{\alpha\overline{\beta}}=\frac{\partial^2 \mathcal{K}}{\partial \phi^\alpha \partial \bar{\phi}^{\bar{\beta}}}$ is the K\"ahler metric, and $\mathcal{K}^{\alpha\overline{\beta}}$ is the inverse of it. $D_\alpha  \mathcal{W} = \frac{\partial  \mathcal{W}}{\partial \phi^\alpha}+\frac{\partial \mathcal{K}}{\partial \phi^\alpha} \mathcal{W}$ is the K\"ahler covariant derivative. 

The factor $e^{\mathcal{K}}$ in the scalar potential leads to the so-called $\eta$ problem in early studies of SUGRA inflation~\cite{Copeland:1994vg}. For any field with canonical K\"ahler  potential $\mathcal{K}=\Bar{\Phi}\Phi$, the second order derivative of the potential:
\begin{equation}
    \frac{\partial^2 V}{\partial \phi \partial \phi} \sim V + \textrm{others}...
    \label{eq:eta_prob}
\end{equation}
leads to a $\mathcal{O}(1)$ slow roll parameter $\eta_V=V''/V \sim 1$. This holds true for the possible inflaton field and any other fields present in the K\"ahler potential. In this paper, we will assume this correction indeed exists for DM field, which means DM gets an effective mass on the order of the Hubble scale during and after inflation. As we will show in this paper, this significantly alters the results for GPP production of DM field.

One elegant solution of the $\eta$ problem is imposing a shift symmetry, which ensures that the inflaton field doesn't appear in the K\"ahler potential. In this way, one can realize most inflation models in ~\cite{Kallosh:2010xz}. 
To embed the aforementioned inflation model in , we use the formalism proposed in~\cite{DallAgata:2014qsj,Kallosh:2014hxa}, which can accommodate an arbitrary inflationary model with arbitrary supersymmetry (SUSY) breaking scale. Our model thus has three superfields: the inflaton superfield $\Phi$, the stabilizer superfield $S$ and the DM superfield $\chi$, with the following K\"ahler potential $\mathcal{K}$ and superpotential $\mathcal{W}$
\begin{equation}
\begin{split}
         \mathcal{K}(\Phi,S,\chi) &= -\frac{1}{2}(\Phi-\Bar{\Phi})^2+S\Bar{S}+\chi\Bar{\chi}\,, \\
        \mathcal{W}(\Phi,S,\chi) &= f(\Phi)+ Sg(\Phi)+\frac{1}{2}m_\chi\chi^2\,.
\end{split}
\end{equation}
We also assume both $\textrm{Im}(\Phi),S,\chi=0$ during inflation as in~\cite{Kallosh:2010xz}. The K\"ahler covariant derivative of each field reads
\begin{equation}
    \begin{split}
        D_\Phi \mathcal{W} &= f'(\Phi)+g'(\Phi) S- (\Phi-\Bar{\Phi})\mathcal{W}\,, \\
        D_S \mathcal{W} &=g(\Phi)+\Bar{S}\mathcal{W}\,,  \\
        D_\chi \mathcal{W}&=m_\chi\chi+\bar{\chi}\mathcal{W}\,.
    \end{split}
\end{equation}
We set $g(\Phi)=\sqrt{3}f(\Phi)$. Up to second order in $\chi$ and ignoring the contribution from $\textrm{Im}(\Phi)$ and $S$, we have from equation \eqref{eq:sugrapotential}
\begin{equation}
\begin{split}
    V_\chi = m_\chi^2\chi \bar{\chi}
    -\frac{1}{2} m_\chi\Bar{\chi}^2f  -\frac{1}{2} m_\chi \chi^2 \Bar{f}
    +\chi \Bar{\chi} \left(f \Bar{f}+f' \Bar{f'} \right)\,,
\end{split}
\end{equation}
where $f' \Bar{f'}= V_I(\phi)$ is the inflaton potential and $\phi=\sqrt{2}\textrm{Re}(\Phi)$ is the actual inflaton.

For $\alpha$ attractor inflation defined in equation~\eqref{eq:t-mode-potential}, the $f$ function has the following form
\begin{equation}
    f(\Phi) = \sqrt{3\alpha V_0} \log\left(\cosh(\frac{\Phi} {\sqrt{3\alpha}})\right)+\frac{m_{3/2}}{\sqrt{3}}\,.
    \label{eq:func_f}
\end{equation}
In the global Minkowski minimum where $\Phi=0$, the SUSY is broken in the $S$ direction. We have:
\begin{equation}
    \abs{D_S \mathcal{W}}^2 = 3 \abs{\mathcal{W}}^2 = m_{3/2}^2\,.
\end{equation}
We can split the complex DM field into two real fields\,, $\chi = (\chi_R + i \chi_I) / \sqrt{2}$\,, with the following mass term:
\begin{subequations}
\begin{align}
    m_{\chi, R}^2 &= m_\chi^2 + V_I + |f|^2 - m_\chi f\,, \\
    m_{\chi, I}^2 &= m_\chi^2 + V_I + |f|^2 + m_\chi f\,,
\end{align}
\end{subequations}
where we have used $f \in \R$ as only $\textrm{Re}(\Phi)$ is non-zero during and just after inflation and there is no mixing between the real and imaginary parts. After the rescaling $\tilde{\chi}_i = a \chi_i$, their effective masses read
\begin{equation}
    \meff^2 = a^2 \left( m_\chi^2  + H^2 + |f|^2 \mp m_\chi f \right).
    \label{eq:m_eff}
\end{equation}
Compared with the effective mass in equation \eqref{eq:mode-eq}, the first distinction is that we have $H^2$ instead of $-a''/a^3$. This is a general feature of SUGRA correction and a result of equation \eqref{eq:eta_prob}, see also \cite{Nakayama:2019yjv}. The last two terms are mostly relevant prior to the oscillation where the inflaton field has large value.

\section{Numerical results}
\label{sec:num}
In this section, we show the necessary numerical results for calculation of DM relic abundance in the current Universe. The first step is to solve the background evolution of inflaton field through equations \eqref{math:background-eq}. In practice, we use the slow-roll solution as initial conditions for the inflaton field. As slow-roll inflation is an attractor, the trajectory converges to the ``true'' one after a sufficient time.

By choosing the above conditions for the background, we also need to ensure that every mode is well inside the horizon upon simulation start and can be described by the Bunch-Davies vacuum
\begin{equation}
    \chi_k = \frac{1}{\sqrt{2 \omega_k}}\,, \; \chi'_k = \frac{-i \omega_k}{\sqrt{2\omega_k}}\,, \quad \text{or equivalently,} \quad \alpha_k = 1, \; \beta_k = 0\,,
\end{equation}
where $\omega_k = \omega_k(\tau_i)$ is taken at the time of initialization $\tau_i$ and it approaches $k$ in the infinite past $\tau_i \rightarrow -\infty$.
Consequently, long wavelength modes need to be initialized early. The largest scale considered here is $k / (a_e H_e) = 10^{-2} $, and it requires initialization at $\sim 10$ e-folds before the end of inflation. 

\begin{figure}[hbt!]
    \centering
    \includegraphics[width=0.9\linewidth]{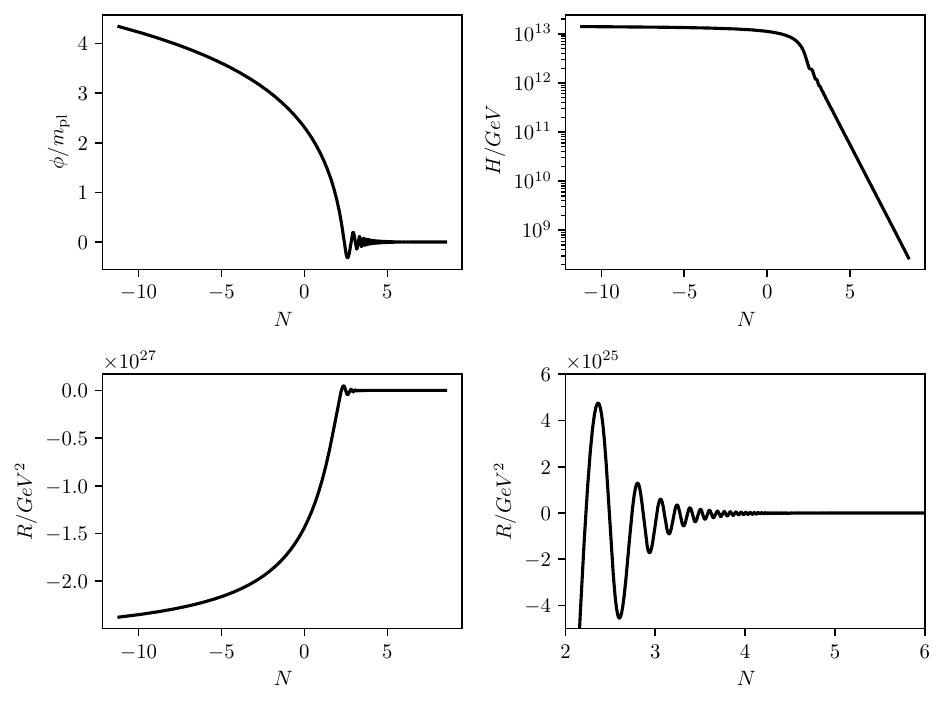}
    \caption{Background evolution: inflaton field value $\phi$, Hubble parameter $H$ and Ricci scalar $R$, with $\alpha \approx \mpl^2$ corresponding to the tensor to scalar ratio $r=0.0035$. The time variable has been transformed to number of $e$-folds $N(t) = \ln(a(t))$.}
    \label{fig:background-plots}
\end{figure}
We show the relevant background quantities in figure \ref{fig:background-plots}. They are computed using the inflation potential given in equation \eqref{eq:t-mode-potential} with a fixed tensor to scalar ratio $r=0.0035$ ($\alpha \approx \mpl^2$).  For a better display of data, the number of $e$-folds $N = \ln a(t)$, instead of the cosmic time $t$, is chosen as the time variable. The scale factor has been rescaled so that the scale factor at end of inflation is unity $a(t_e) \equiv 1$, which was determined by the condition that the first Hubble slow-roll parameter equals $0.1$:
\begin{equation}
    \epsilon_1 := - \frac{\dot{H}}{H^2} = \frac{1}{2\mpl^2} \left( \dv{\phi}{N} \right)^2 \stackrel{!}{=} 0.1\,.
\end{equation}
Here we use dot to denote derivative with respect to the cosmic time $t$ and the second equality follows from the Friedmann equations \eqref{math:background-eq}.
Thus, the number of $e$-folds at the end of inflation is defined as zero. The first diagram shows the evolution of the field value $\phi(N)$. It smoothly rolls down to the minimum at $\phi=0$ and starts to oscillate around it. This oscillation process induces the decrease of Hubble scale $H$ as well as an oscillation of the Ricci scalar $R$. 
During slow roll inflation, both two quantities are more or less constants:
\begin{equation}\label{math:ricci_inf}
    R = - 6 \frac{a''}{a^3} \stackrel{\text{SR}}{\approx} -12 H^2 \approx \text{constant}\,.
\end{equation}
However, the Ricci scalar oscillates around zero after inflation, while $H$ decays exponentially and slightly oscillates. Note that the Ricci scalar $R$ appears in the effective $\chi$ mass for non-SUGRA case in equation \eqref{eq:mode-eq}, and the (square of) Hubble parameter $H^2$ for the SUGRA case in equation \eqref{eq:m_eff}.

Let us first consider the effective frequency squared without SUGRA correction; it takes the following form during slow-roll inflation
\begin{equation}
    \omega_k^2 
    = k^2 + a^2 \left( m_\chi^2 + \frac{R}{6} \right)\stackrel{\text{SR}}{=} k^2 + a^2 \left( m_\chi^2 - \frac{2 \mpl^2}{\alpha} m_\phi^2   \right)\,,
    \label{eq:omega_nosugra}
\end{equation}
where the second Friedman equation in equations \eqref{math:ricci_inf} have been used. The last equality only holds during slow-roll inflation. For light DM field, the oscillation of $R$ leads $\omega_k^2$ to cross zero and thus induces a tachyonic instability, especially for low momentum modes. 
\begin{figure}[hbt!]
    \centering
    \includegraphics[width=0.6\textwidth]{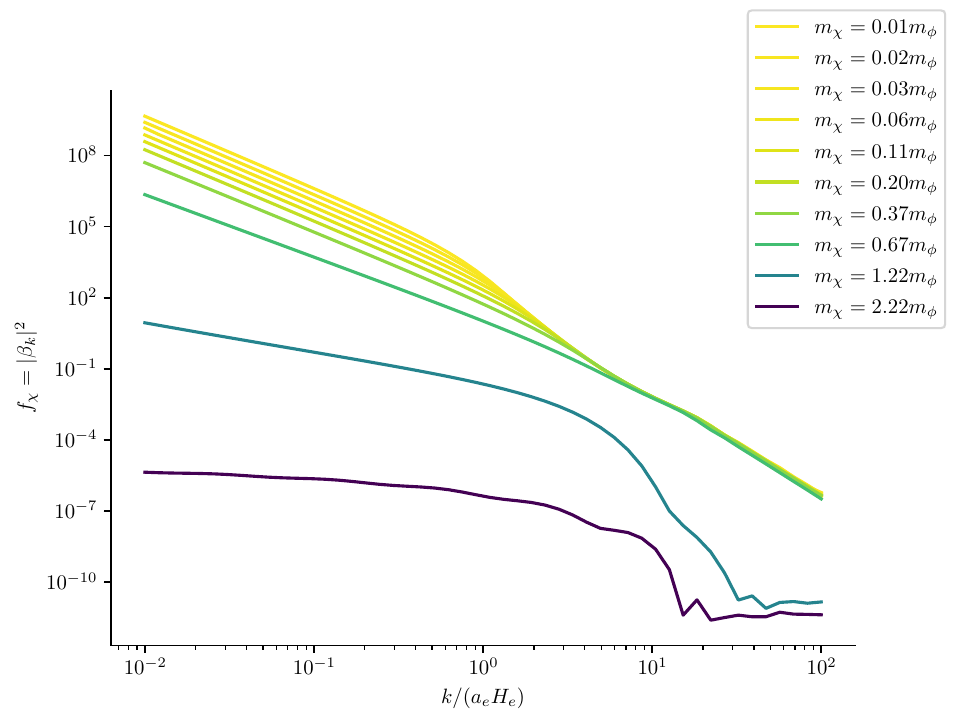}
    \caption{Phase space distribution of $\chi$ quanta without SUGRA correction. In the vanishing mass limit, the phase space distribution approaches $|\beta_k|^2 \propto k^{-3}$.}
    \label{fig:f_nosugra}
\end{figure}

One can then solve the differential equation associated with $\omega_k$ to get the phase space distribution of the DM field. It proves to be more efficient to use equation \eqref{eq:mode-eq} due to the tachyonic instability. We show some examples with different DM mass in figure \ref{fig:f_nosugra}. For light fields, the effect of the tachyonic instability is significant, which leads to the typical scaling behavior in the low momentum region $f_\chi \propto k^{3}$. In contrast, for heavier DM field, $f_\chi$ approaches a constant for small momentum. The threshold DM mass is roughly the inflaton mass, as expected from equation \eqref{eq:omega_nosugra}. Heavier DM fields are generally harder to get excited. Their $f_\chi$ are much smaller than those of light fields. 

\begin{figure}
    \centering
    \includegraphics[width=0.5\textwidth]{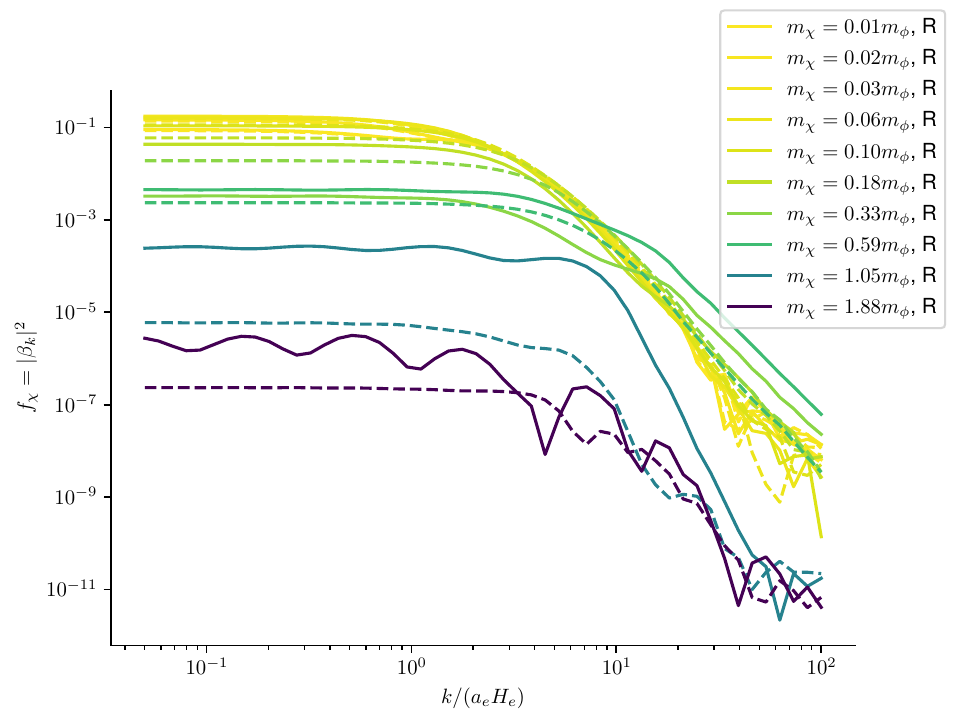}%
    \includegraphics[width=0.5\textwidth]{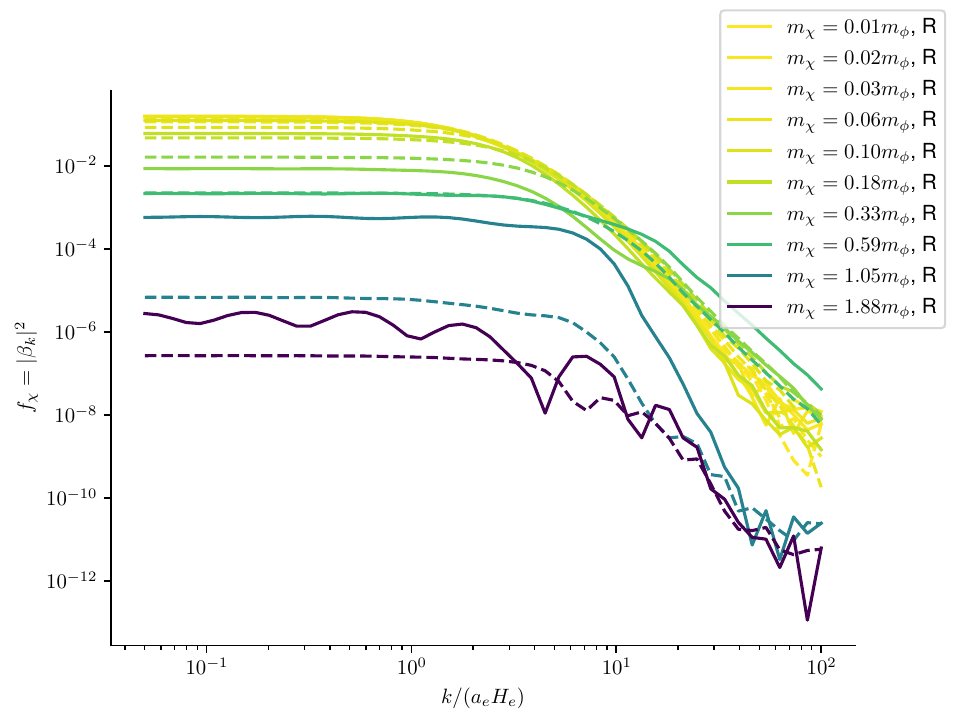}
    \caption{Phase space distribution of $\chi$ quanta with $m_{3/2}=0$ (left) and $m_{3/2}= 0.1 m_\phi$ (right). Dashed line represents the imaginary field and solid line the real field.}
    \label{fig:f_0.0035}
\end{figure}

Now let us consider the GPP of DM particles with SUGRA correction. The spectra for two distinct values of $m_{3/2}$ are shown in figure \ref{fig:f_0.0035}.  The most striking feature is that the spectrum has no infrared divergence. Due to the additional terms in the effective mass in equation \eqref{eq:m_eff}, the tachyonic instability doesn't exist in SUGRA case. Again, higher $m_\chi$ gives less $\chi$ particles. When $m_\chi \ll m_{3/2}$, the spectra is controlled by the inflation potential and size of $m_{3/2}$. All light fields share basically indistinguishable spectra. 

\begin{figure}
    \centering
    \includegraphics[width=0.5\linewidth]{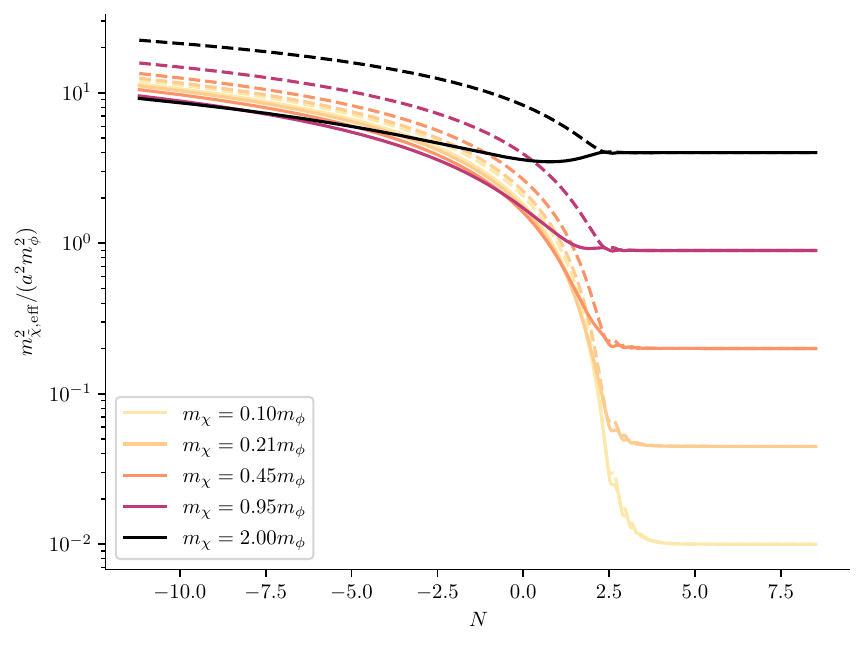}%
    \includegraphics[width=0.5\linewidth]{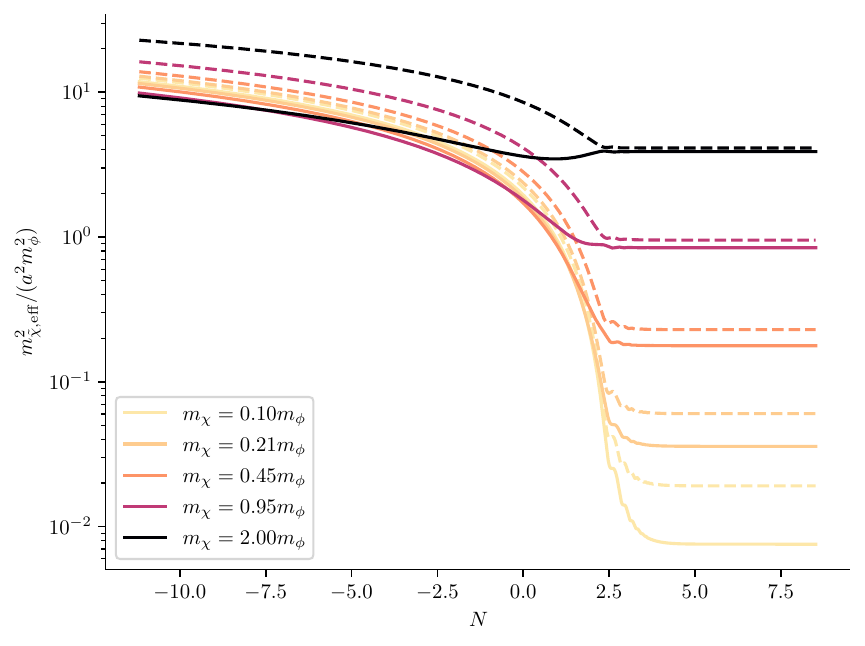}
    \caption{Effective mass squared in the appropriate units with $\mgrav = 0.0$ (left) and $\mgrav = 0.1 m_\phi$  (right) respectively. Solid (dashed) line is for effective mass of real (imaginary) field.}
    \label{fig:0.0035_m_eff}
\end{figure}

It is beyond the scope of the current work to have an analytic understanding of the particle number spectra. For a qualitative understanding, the time evolution of the effective mass might give some insights. We show the effective mass of DM fields during and after inflation in figure \ref{fig:0.0035_m_eff}. The imaginary field has higher mass during inflation and the rate of change in effective frequency is higher for $m_\chi < m_\phi$. Thus, as one can see from the figure \ref{fig:f_0.0035}, in SUGRA the imaginary part has higher production in such case. On the contrary, for $m_\chi \gtrsim m_\phi$, the real part is produced in greater amount. When $m_{3/2} \ll \alpha m_\phi / \mpl^2 $, the mass splittings prior to reheating are similar among $m_{3/2}$'s, as the first term in the $f$-function dominates, see equation \eqref{eq:func_f}. Although the splitting at the end of the particle production clearly depends on the gravitino mass
\begin{equation}
    m_{\chi, R,I}^2 = m_\chi^2 + \frac{m_{3/2}^2}{3} \mp \frac{m_{3/2} m_\chi}{\sqrt{3}}\,,
\end{equation}
the ultimate rate of change for $m_{\chi, \text{eff}}^2$ and $\omega_k^2$ are largely unaffected by it. Thus, the phase space distributions are very similar with light gravitino masses.

\begin{figure}[b]
\centering
\begin{subfigure}{0.48\textwidth}
    \includegraphics[width=\textwidth]{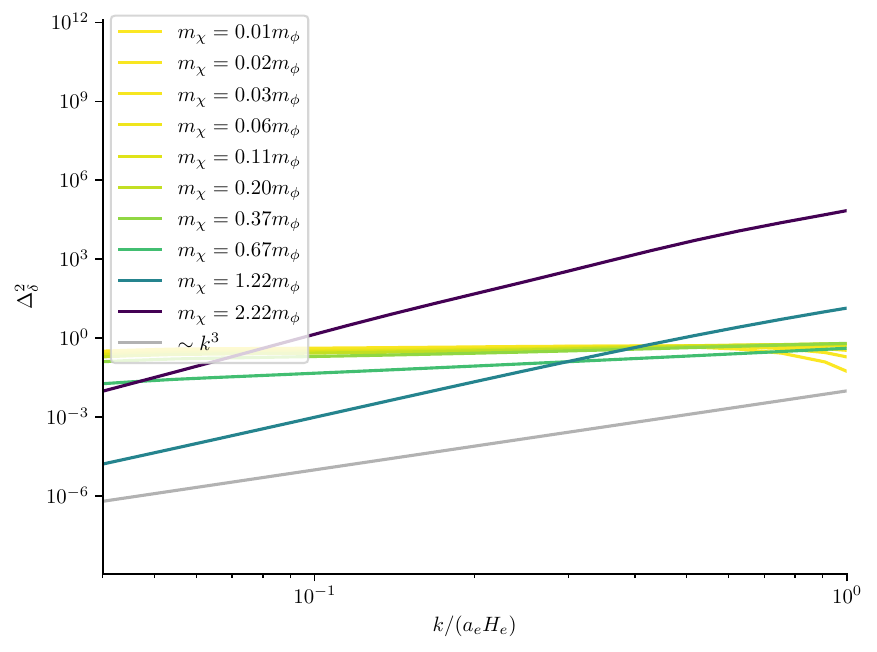}
    \caption{Without SUGRA correction. Computed from the phase space distribution shown in figure \ref{fig:f_nosugra}.}
    \label{fig:iso-nosugra}
\end{subfigure}
\hfill
\begin{subfigure}{0.48\textwidth}
    \includegraphics[width=\textwidth]{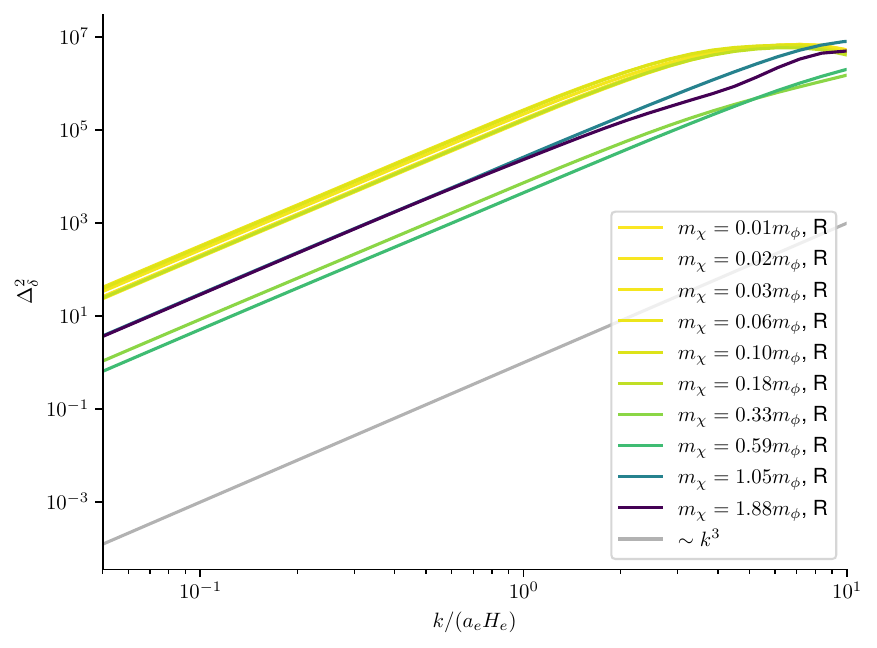}
    \caption{With SUGRA correction and vanishing $m_{3/2}$. Computed from the phase space distribution shown in figure \ref{fig:f_0.0035}. Only the real field contributions are shown here.}
    \label{fig:iso-sugra}
\end{subfigure}
\caption{Isocurvature power spectrum calculated using equation \eqref{eq:iso_formula}.}
\label{fig:isocurv}
\end{figure}

From the phase space distribution, the isocurvature spectrum can be calculated via equation~\eqref{eq:iso_formula}. It has been shown (semi-)analytically that $|\beta_k|^2 \propto k^{-3}$ (in the low momentum side) produces a nearly scale-invariant isocurvature spectrum in the low momentum side and $|\beta_k|^2 \propto \text{const.}$ gives blue-tilted spectrum \cite{Kolb_2023}. In the vanishing mass limit $m_\chi \rightarrow 0$, the isocurvature spectrum indeed approaches scale invariance, see figure \ref{fig:iso-nosugra}. This can be potentially dangerous when considering CMB isocurvature constraints~\cite{Ling:2021zlj}. In the SUGRA case, the isocurvature spectrum always holds a $k^3$ scaling in the low momentum limit. The spectra are shown in figure \ref{fig:iso-sugra}. Thus, the CMB isocurvature constraint poses no limitation on the model parameters, when SUGRA corrections are present.

\begin{figure}
    \centering
    \includegraphics[width=0.5\textwidth]{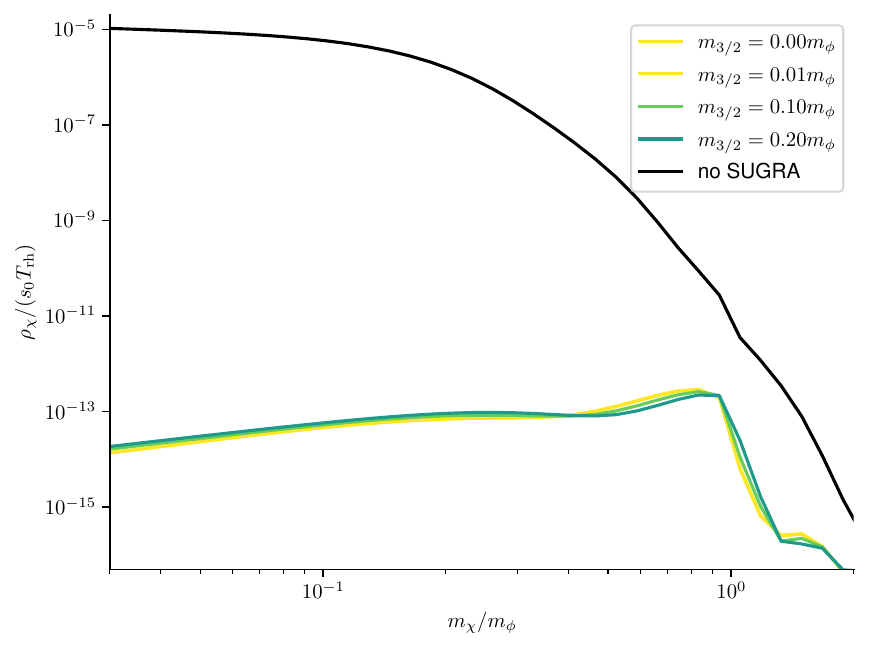}%
    \caption{Comoving energy density of the DM field $\chi$ divided by the reheating temperature. Higher values require lower reheating temperatures. The tensor-to-scalar ratio is fixed to $r=0.0035$.}
    \label{fig:TMode-abundance}
\end{figure}

We show the energy densities of DM in the current universe in figure \ref{fig:TMode-abundance}. We assume that the DM particle does not interact with other (standard model) particles and their comoving number density is conserved after production. Thus, its current energy density can be expressed as $\rho_{\chi, 0} \approx m_\chi n_{\chi,0}$. On the vertical axis, we use the energy density of the DM particles divided by the entropy density and reheating temperature $\rho_{\chi, 0} / (s_0 T_{\text{rh}}) $. This quantity is obtained from comoving number via the equation \eqref{eq:rho_s0}. To get the desired relic abundance $\Omega_{\textrm{DM}} h^2 \approx 0.12$, we need
\begin{equation}
    \frac{\rho_{\chi,0}}{s_0} \approx \SI{4e-10}{\giga \electronvolt}\,.
    \label{eq:rho_s0}
\end{equation}
This means different $m_\chi$'s will need different reheating temperature to satisfy this condition. 

The phase space distribution of non-SUGRA case scales like $k^{-3}$ in the infra-red limit. This leads to an IR divergence in the total number density equation \eqref{eq:number-density} with $n \sim  \int_0^\infty \dd{k} k^{-1}$. To cure this divergence, one can set an IR cutoff at CMB scale, since DM perturbations larger than the horizon at recombination are just part of the background. As it is expensive to track modes across so many number of $e$-folds in our numerical computation, we choose to extrapolate the phase space distribution to the CMB scale. We take $60$ $e$-folds of inflation since CMB scale. It can be shown that the final abundance is proportional to the number of $e$-folds, if the number of $e$-folds is chosen differently. The abundance without SUGRA correction can be found in figure \ref{fig:TMode-abundance}. As we see, the abundance saturates at the low mass limit, as expected from the phase space distribution in figures \ref{fig:f_nosugra}. The tachyonic instability goes away with increasing $m_\chi$, therefore, the abundance decreases rapidly.

The number density with SUGRA correction is more straightforward to compute and it is shown in figure \ref{fig:TMode-abundance} with $r=0.0035$. When $m_\chi < m_\phi$, the abundance from GPP increases with $m_\chi$ linearly, which comes from the linear mass dependency in energy density of non-relativistic particles. The abundance then decreases exponentially when $m_\chi > m_\phi$. There is an interesting bump for small $m_{3/2}$ around $m_\chi \sim m_\phi$. This primarily comes from the real field contribution, and one can see this for $k/(a_e m_\phi) \gtrapprox 1$ also in figure \ref{fig:f_0.0035}.

\begin{figure}
    \centering
    \includegraphics[width=0.5\textwidth]{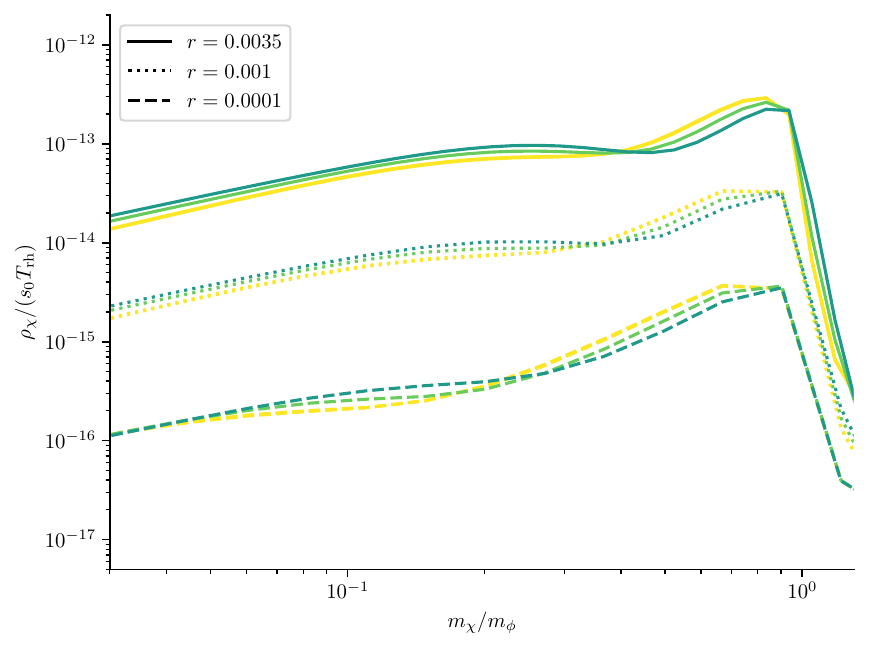}%
    \includegraphics[width=0.5\textwidth]{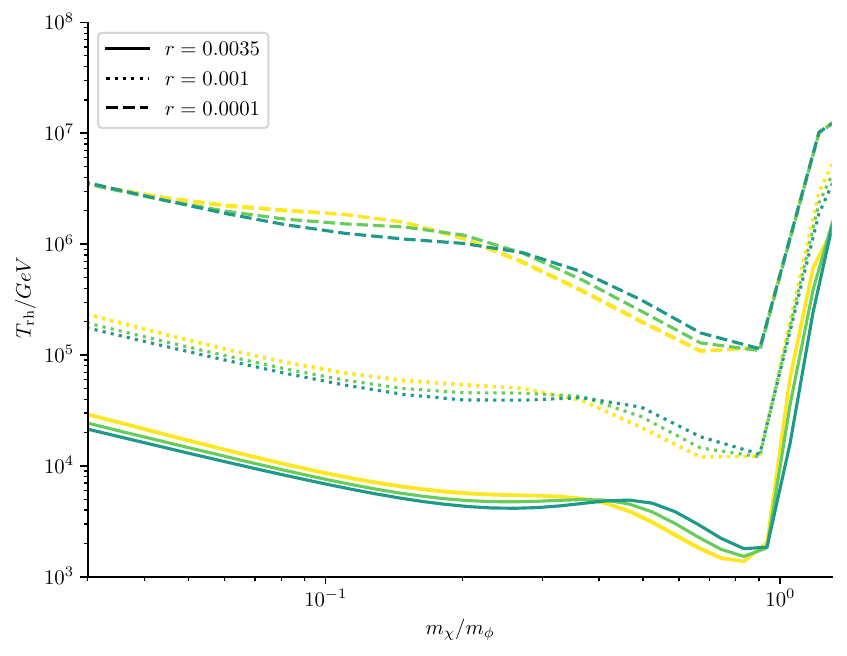}
    \caption{DM abundance $\rho_{\chi, 0}/(s_0 T_\text{rh})$ and required reheating temperature for $\chi$ to be the only DM. Different colors correspond to different $m_{3/2}$, c.f. figure \ref{fig:TMode-abundance}.}
    \label{fig:abund-diff-r}
\end{figure}

The abundances with various desired values of tensor-to-scalar ratio are plotted in figure \ref{fig:abund-diff-r}. As we wish $\chi$ to be the only DM around, equation \eqref{eq:rho_s0} leads to certain reheating temperature for a set of fixed model parameters ($m_{3/2}$, $m_\chi$ and $r$). In SUGRA a high reheating temperature may lead to the gravitino problem~\cite{Khlopov:1984pf, Ellis:1984eq, Kawasaki:1994af}. If $m_{3/2} \approx 1 \textrm{TeV}$, this means  an upper bound on reheating temperature $T_\text{rh} < 10^9 \textrm{GeV}$. Our results are compatible with this bound.  As implied in figure \ref{fig:TMode-abundance}, required reheating temperature $T_\text{rh}$ for all values of $r$ goes like $m_\chi^{-1}$ for small $m_\chi$ until $m_\chi \sim m_\phi$, where the produced comoving number density drops dramatically and high reheating temperature is necessary. Effects of $m_{3/2}$ is still very minimal. Lower desired tensor-to-scalar ratio reduces the Hubble scale $H_e$, suppresses the GPP production and demands higher reheating temperature.

\clearpage
\section{Conclusion and outlook}
\label{sec:conc}
In this work, we looked at gravitationally produced DM particles with SUGRA corrections. The $\alpha$-attractor inflation model was chosen for this investigation. It is theoretically well-motivated and fits the observation well. We solve the equation governing Fourier modes of the DM field, equation \eqref{eq:mode-eq}, on top of the evolving background. Then, the Bogoliubov coefficients are reconstructed and (its modulus square) can be interpreted as particle number spectrum. In the vanilla scenario without SUGRA corrections, the produced particle number decrease exponentially once the DM particles are roughly heavier than the inflaton and there is clear tachyonic instability. However, the particle production with SUGRA correction has some distinctive features. The phase space distribution shows no infra-red divergence due to the lack of tachyonic instability. Computation of the current abundance, thus, requires no cut-off. The isocurvature power spectrum is blue-titled ($\propto k^3$) and not constrained by CMB isocurvature non-detection. The required reheating temperature is shown to increase with decreasing tensor-to-scalar ratio $r$, and higher than non SUGRA cases.

There are a couple of directions for further investigation. We focus only on the $\alpha$-attractor potential with $n=1$. One can then ask how different assumptions would change our results. The potential can be written as $V \propto |\phi|^{2n}$ near the origin. This changes the inflaton mass (there is no quadratic term in potential with $n \geq 2$) and the equation of state during the reheating phase changes. The formula \eqref{eq:rho_s0} is derived assuming matter domination for reheating. Additionally, there are non-linear phenomena in the inflaton sector, which can further change the background evolution, see e.g. \cite{lozanovSelfresonanceInflationOscillons2018} and the review \cite{aminNonperturbativeDynamicsReheating2015}. In this work, the results are all computed numerically. One might gain additional insight by deriving analytical results. How to incorporate non-minimal coupling in SUGRA setting would be an interesting direction. Introduction of a non-minimal coupling to the scalar field famously introduces massively different spectrum. It would be interesting to see such effects with SUGRA correction as well. We leave these for further works.

The $\chi$ field does not have to be the DM; instead there are several other possibilities. GPP can also be used as a portal to reheating the hidden sector \cite{Barman_2022}. GPP could also be an important channel to produce the observed baryon-asymmetry, where heavy $X$ particles are produced gravitationally, where decay violates baryon number and $CP$ conservation \cite{flores2024rolecosmologicalgravitationalparticle}.

GPP can also be applied to other fields with higher spin. The spin-$1/2$ superpartner of $\chi$ field will be produced on an equal footing. As the expansion of the universe can also break SUSY characterized by the Hubble parameter~\cite{Dine:1983ys,Bertolami:1987xb,Copeland:1994vg,Dvali:1995mj,Dine:1995uk}, the mass splitting between scalar and its fermionic superpartner becomes time dependent, which may lead to novel result. Another important example is the tensor modes (spin-$2$ field) in the metric perturbations which (can) corresponds to the gravitational waves. Gravitational waves are inevitably produced via the same mechanism as the scalar field: they have the same dispersion relation as massless non-minimal scalar field (at least in the non-SUGRA case). The only difference is that tensor mode has two polarizations. They would correspond to frequencies from radiation-matter equality scale $f_\text{eq} \sim \SI{5e-17}{\hertz}$ to $f_{\text{rh}} \sim \SI{200}{\hertz} \left( \frac{ T_{\text{rh} }}{10^9 \si{\giga \electronvolt} } \right)$ \cite{Kolb:2023ydq}. As the required reheating temperature is different between SUGRA and non-SUSY case, these frequencies shifts to $\sim \SI{1}{\hertz} - 10^{-4} \si{\hertz}$ in our model. It is well within accessible frequency ranges of e.g. LISA. However, the energy density parameter would be $\Omega_{\text{gw}, 0} h^2 \sim 10^{-15}$ below LISA sensitivity \cite{Moore_2014}.

\clearpage
\appendix
\section{Calculation for DM energy density to entropy density ratio}
\label{sec:app_rho_s0}
\begin{figure}[hbt]
    \centering
    \includegraphics[width=0.9\linewidth]{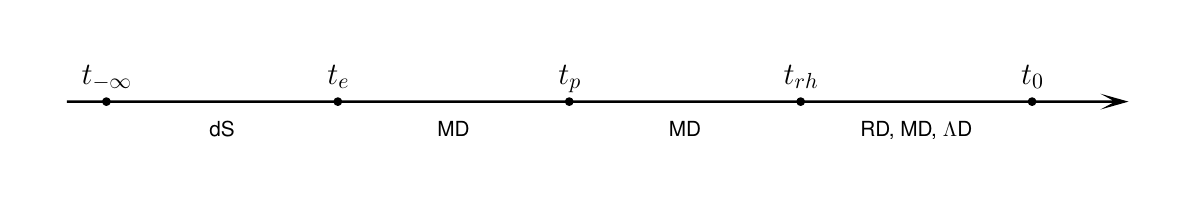}
    \caption{Thermal history of our model. Here $t_{-\infty}$ represents the initial time in our simulation. $t_{e}$ represents the time when inflation has ended and inflaton starts to oscillates around its minimum. $t_p$ denotes the time when GPP is almost finished and $t_{\textrm{rh}}$ denotes the time when reheating is complete. We also use $t_0$ to denote the current time.}
    \label{fig:thermalhistory}
\end{figure}
In this section we briefly review the thermal history of our universe and explain how we calculate the density to entropy ratio explicitly. Our simulation starts when the inflaton $\phi$ field stays in the inflation region, which we will formally denote it by $t_{-\infty}$. At $t=t_{e}$  the (Hubble) slow roll conditions are broken and $\phi$ field start to oscillate around its minimum. In this paper we consider the case where the oscillation part can be approximated by $\phi^2$, which means the energy density of the inflaton field red-shift as matter. 
This transition from de Sitter spacetime to matter dominated space also triggers the gravitational production of the super heavy DM.
Most of the super heavy DM are produced at the beginning of the oscillation part, and we assume at $t=t_p$ this production is basically complete and its energy density also red-shifts as matter. Hence we have the following identity
\begin{equation}
    \frac{\rho_\chi(t_p)}{\rho_\phi(t_p)}=\frac{\rho_\chi(t_{\textrm{rh}})}{\rho_\phi(t_{\textrm{rh}})}=\frac{\rho_\chi(t_{\textrm{rh}})}{\rho_r(t_{\textrm{rh}})},
\end{equation}
where we have used that DM $\chi$ field redshifts in the same way as the inflaton $\phi$ field, and we define the reheating temperature when $\rho_\phi(t_{\textrm{rh}})=\rho_r(t_{\textrm{rh}})$. Here $\rho_r$ denotes the energy density in radiation. After the reheating is complete, we have
\begin{equation}
    \frac{\rho_\chi(t_{\textrm{rh}})}{\rho_r(t_{\textrm{rh}})}=\frac{\rho_\chi(t_0)}{\rho_r(t_0)}\frac{T_0}{T_\textrm{rh}}.
\end{equation}
The entropy density and energy density in today's Universe reads
\begin{equation}
\begin{split}
    s_0&=g_*\frac{2\pi^2}{45}T_0^3,\\
    \rho_{r,0}&=g_*\frac{\pi^2}{30}T_0^4.
\end{split}
\end{equation}
Hence, we can rewrite $\rho_{r,0}=\frac{3}{4}sT_0$ and the energy density to entropy density reads
\begin{equation}
\frac{\rho_\chi(t_0)}{s_0}=\frac{3}{4}T_0\frac{\rho_\chi(t_0)}{\rho_r(t_0)}=\frac{3 m_\chi T_{\textrm{rh}}}{4}\frac{n_\chi(t_p)}{\rho_\phi(t_p)},
\end{equation}
where we have used $\rho=m n$ for non-relativistic particles (CDM). The energy density of $\rho_\phi(t_p)$ can be connected to the energy density of inflaton at the end of inflation
\begin{equation}
    \rho_\phi(t_p)=\rho_{\phi} (t_{e})\left(\frac{a({t_e})}{a(t_p)}\right)^3=3\mpl^2 H^2(t_{e})\left(\frac{a({t_e})}{a(t_p)}\right)^3.
\end{equation}
The finial expression reads
\begin{equation}
    \frac{\rho_\chi(t_0)}{s_0} =\frac{ m_\chi T_{\textrm{rh}} H(t_e)}{4\mpl^2}\left(\frac{a^3(t_p) n_\chi(t_p)}{a^3({t_e}) H^3(t_e)}\right) .
    \label{eq:rho_s0}
\end{equation}
Current observation suggests this quantity should be equal to $\SI{4e-10 }{\giga \electronvolt}$ to contribute the full amount of DM.

\acknowledgments
We thank Manuel Drees for useful discussion and commenting on the draft. We thank the package developers of \cite{Hunter:2007, 2020SciPy-NMeth, harris2020array, rackauckas2017differentialequations}.

\newpage
\bibliographystyle{JHEP}
\bibliography{main}
\end{document}